\documentclass[aps,prb,twocolumn,showpacs]{revtex4}
\newcommand{\bsim}{\mbox{\raisebox{-0.1cm}{$\;
\stackrel{\textstyle>}{\sim}\;$}}}
\newcommand{\lsim}{\mbox{\raisebox{-0.1cm}{$\;
\stackrel{\textstyle<}{\sim}\;$}}}
\usepackage{psfig}

\begin{document}

\title{Charge fluctuations and electron-phonon interaction
in the finite-$U$ Hubbard model}

\author{E. Cappelluti$^{1,2}$,
B. Cerruti$^2$ and L. Pietronero$^2$}

\affiliation{$^1$``Enrico Fermi'' Research Center, c/o Compendio del Viminale,
v. Panisperna 89/a, 00184 Roma, Italy}

\affiliation{$^2$Dipart. di Fisica, Universit\`{a} di Roma ``La Sapienza",
P.le A.  Moro, 2, 00185 Roma,
and INFM UdR Roma1, Italy}

\date{\today}

\begin{abstract}
In this paper we employ a gaussian expansion within the finite-$U$
slave-bosons formalism to
investigate the momentum structure of the electron-phonon 
vertex function in the Hubbard model as function of $U$ and $n$.
The suppression of large momentum scattering and the onset a 
small-${\bf q}$ peak structure, parametrized
by a cut-off $q_c$,
are shown to be essentially ruled
by the band narrowing factor $Z_{\rm MF}$ due to the
electronic correlation. A phase diagram of $Z_{\rm MF}$ and $q_c$ in the
whole $U$-$n$ space is presented.
Our results are in more than qualitative agreement
with a recent numerical analysis and
permit to understand some anomalous features
of the Quantum Monte Carlo data.
\end{abstract}
\maketitle

In the past months a variety
of experiments have pointed out
an important role of the electron-phonon (el-ph) interaction in many physical
properties of the
cuprates.\cite{lanzara,keller,dastuto}
These recent findings have triggered
a renewed interest for a theoretical understanding of the el-ph
properties in strongly correlated systems.

One of the most remarkable effects of the strong electronic correlation
on the el-ph properties is a to favor
forward (small ${\bf q}$) scattering
in the electron-phonon vertex, ${\bf q}$ being the exchanged
phonon momentum.
This feature was investigated in the past
by means of analytical techniques based on slave-bosons
or $X$-operators.\cite{kim,grilli,zeyher,kulic}
The assumption of forward scattering predominance within an el-ph
framework was shown to explain in a natural way
several anomalous properties of cuprates as the difference between transport
and superconducting el-ph coupling constants,\cite{zeyher}
the linear temperature behaviour of the resistivity,\cite{cp}
the $d$-wave symmetry of the superconducting gap.\cite{kulic,pgp}
Small ${\bf q}$ scattering selection was shown moreover to be
responsible in a natural way for high-$T_c$ superconductivity
within the context of the nonadiabatic superconductivity.\cite{gpsprl}

So far, this important feature was analyzed only by means of
analytical approaches in the $U = \infty$ limit
and a definitive confirm of it based
on numerical methods was lacking. 
The charge response at finite $U$ was addressed
in Refs. \onlinecite{lavagna,kaga},
which however focused only on ${\bf q}=0$ susceptibilities.
With these motivations
in a recent paper Huang {\em et al} have addressed this issue
in the twodimensional Hubbard model with generic $U$
by using Quantum Monte Carlo (QMC) techniques
on a $8 \times 8$ cluster.\cite{huang}
Their results provide a good agreement with the previous analytical
studies and represent an important contribution to assess the relevance
of el-ph interaction in correlated system.

In this paper we employ the slave-boson techniques to investigate
the evaluation of the momentum structure of the el-ph vertex interaction
as function of the Hubbard repulsion $U$.
While the previous analytical studies
were limited to the $U = \infty$ limit,\cite{kim,grilli,zeyher,kulic}
we were able in this way
to evaluate the small momenta
selection in the whole phase diagram of parameters $U$-$n$,
where $n$ is the electron filling.
In particular we show that the predominance of
small ${\bf q}$ scattering is strongly dependent on the closeness
of the system to the metal-insulator transition (MIT). Our analysis
suggests that forward scattering is essentially ruled by the one-particle
(mean-field) parameter $Z_{\rm MF}$ which represents the band narrowing factor
due to the correlation effects. We also discuss some anomalous features
of the numerical results of Ref. \onlinecite{huang}. The fair agreement between
our slave-boson results and the Quantum Monte Carlo data \cite{huang}
supports the reliability of our approach to investigate these issues.

Working tool of this paper will be the four slave-boson method first
introduces by Kotliar and Ruckenstein to investigate the Hubbard model
at finite $U$. In this approach the physical Hilbert space is enlarged
by introducing, in addition to the standard fermionic degrees of freedom
$c_{i\sigma}$, four auxiliary bosons, $e$, $d$ $p_\sigma$,  which count
respectively empty, double and single occupied (with spin $\sigma$)
sites.\cite{kotliar}
The extended Hilbert space is thus restricted to the physical one
by the Lagrange multipliers $\lambda^{(1)}$, $\lambda^{(2)}_\sigma$.
The Hubbard Hamiltonian reads thus:
\begin{eqnarray}
H &=& \sum_{ij \sigma} Z_{ij\sigma}[e,d,p_\sigma,p_{-\sigma}] t_{ij}
c_{i\sigma}^\dagger c_{j\sigma}
\nonumber\\
&&+\sum_i \lambda_i^{(1)} 
\left[ e_i^\dagger e_i +\sum_\sigma p_{i\sigma}^\dagger p_{i\sigma}
+d_i^\dagger d_i-1\right]
\nonumber\\
&&+\sum_{i\sigma} \lambda_{i\sigma}^{(2)}
\left[ c_{i\sigma}^\dagger c_{i\sigma}
-p_{i\sigma}^\dagger p_{i\sigma} -d_i^\dagger d_i  \right],
\label{hubbard}
\end{eqnarray}
where $t_{ij}$ is the bare hopping term, here considered
within a nearest neighbor model, and $Z[\ldots]$ is the reduction of
the hopping term due to the correlation effects.
A standard method to account for charge, spin, and different
kinds of fluctuations is to expand Eq. (\ref{hubbard}) at the
gaussian level around its mean-field solution for the 
auxiliary fields $e$, $d$, $p_\sigma$, $\lambda^{(1)}$,
$\lambda^{(2)}_\sigma$.\cite{lavagna,becca}
Eq. (\ref{hubbard}) can be thus re-written as:
\begin{eqnarray}
H&=& \sum_{{\bf k} \sigma} \left[Z_{\rm MF} t_{\bf k} 
+\lambda^{(2)}_\sigma \right]
c_{{\bf k}\sigma}^\dagger c_{{\bf k}\sigma}
+\sum_{{\bf q} \mu}
\alpha_{{\bf -q}}^\mu
\left[ B^{-1}_{\bf q} \right]^{\mu \nu}
\alpha_{{\bf q}}^\nu
\nonumber\\
&&+\sum_{{\bf k} {\bf q} \sigma \mu}
\Lambda_{{\bf k},{\bf q}}^\mu
c_{{\bf k+q}\sigma}^\dagger c_{{\bf k}\sigma} \alpha_{{\bf q}}^\mu
+ \mbox{const.},
\label{gauss}
\end{eqnarray}
where the first term describes the mean-field approximation where
all the auxiliary fields are determined by their self-consistent
saddle point solution, the second term are the quadratic fluctuations
of the auxiliary fields
about their mean-field solution, comprehensive of the electronic loop
self-energy contributions,\cite{lavagna,becca}
and the third one is the linear electron-slave-boson interaction.
The index $\mu$, $\nu$  run on the seven auxiliary fields.

The el-ph vertex function $g({\bf k},{\bf q},\omega)$
in the presence of electronic correlation can be promptly determined
as the linear response of the systems described by Eq. (\ref{gauss})
to an external charge excitation. In the paramagnetic case
the spin index $\sigma$ can be dropped and the auxiliary field space
reduce to five components. At the gaussian level we have \cite{lavagna,becca}
\begin{equation}
g({\bf k},{\bf q};\omega) =1 - 2 \sum_{\mu \nu}
\Lambda_{{\bf k},{\bf q}}^\mu 
\left[ D^{-1}_{\bf q}(\omega) \right]^{\mu \nu}
\tilde{\Pi}^\nu_{\bf q}(\omega),
\label{vertex}
\end{equation}
where the factor 2 takes into account the spin degeneracy,
$D^{-1}$ is the renormalized slave-boson propagator,
$2\left[D^{-1}_{\bf q}(\omega)\right]^{\mu \nu}
= 2\left[B^{-1}_{\bf q}\right]^{\mu \nu} + \Pi^{\mu \nu}_{\bf q}(\omega)$,
which includes
the ``bubble'' contribution
to the slave-boson self-energy
\begin{equation}
\Pi^{\mu \nu}_{\bf q}(\omega) =
\sum_{\bf k} 
\Lambda_{{\bf k},{\bf q}}^\mu \Lambda_{{\bf k},{\bf q}}^\nu 
\frac{f(E_{\bf k})-f(E_{\bf k+q})}{E_{\bf k+q}-E_{\bf k}-\omega},
\label{doublePi}
\end{equation}
and where
\begin{equation}
\tilde{\Pi}^\nu_{\bf q}(\omega) =
\sum_{\bf k} 
\Lambda_{{\bf k},{\bf q}}^\nu 
\frac{f(E_{\bf k})-f(E_{\bf k+q})}{E_{\bf k+q}-E_{\bf k}-\omega}.
\label{tildePi}
\end{equation}
Here $f(x)$ is the Fermi factor and $E_{\bf k}$ is the renormalized
electronic dispersion $E_{\bf k}= Z_{\rm MF} t_{\bf k}$.

We are now in the position to apply Eqs. (\ref{vertex})-(\ref{tildePi})
to calculate the ${\bf q}$-structure of the electron-phonon
vertex function in the presence of electronic correlation
for generic Hubbard repulsion $U$ and electron filling $n$.
From the technical point of view we use a slightly different
version of the four slave-boson technique as introduced 
by Kotliar and Ruckentstein. We note indeed that Eq. (\ref{gauss}),
when derived in the context of the Kotliar-Ruckenstein formalism,
present serious pathologies at finite $U$ in the limit $n \rightarrow 0,2$
where the linear electron-slave-boson matrix elements
diverge.\cite{cerruti}
As pointed out in Ref. \onlinecite{kotliar} however
several choices for the auxiliary fields $e$, $d$, $p_{i\sigma}$
can be employed as long as they fulfill the constraints imposed
by the real Hilbert space. We make use of this freedom to avoid
those unphysical divergences. The analytical derivation of
Eq. (\ref{gauss}) in this new basis of auxiliary fields is long and
cumbersome, but quite straightforward and it does not affect
in a significant way our results which focus on the MIT transition
Technical details will be thus presented in a longer publication.

\begin{figure}
\centerline{\psfig{figure=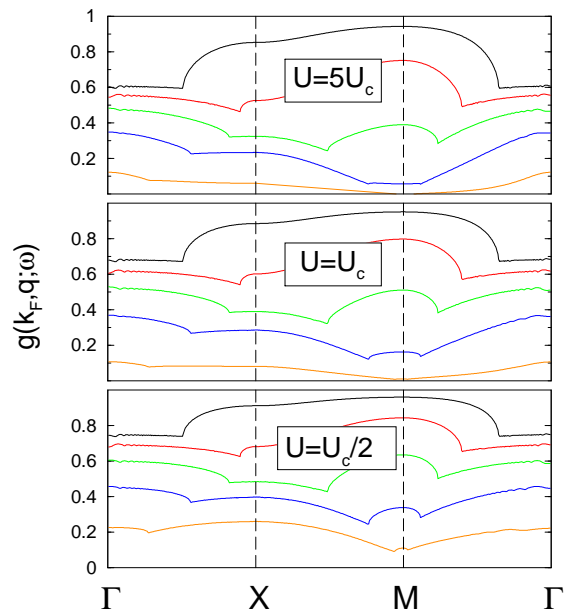,width=7.5cm,clip=}}
\caption{(color online) Momentum dependence
of the electron-phonon vertex function
$g(k_{\rm F},{\bf q};\omega=0)$ 
in the presence of electronic correlation for different $U/U_c$ and
$n$. In each panel solid lines represent, from the top to the bottom:
$n=0.1,0.3,0.5,0.7,0.9$.  Standard labels for the
high symmetry points correspond to $\Gamma=(0,0)$, $X=(0,\pi)$,
$M=(\pi,\pi)$.}
\label{g-vs-q-u}
\end{figure}
In Fig. \ref{g-vs-q-u}
we plot the evolution of momentum structure of the
electron phonon vertex function $g(k_{\rm F},{\bf q};\omega=0)$ 
for $\omega=0$ and
${\bf k}=k_{\rm F}$
lying on the Fermi surface for different electron filling $n$
and $U$.
The top panel in Fig. \ref{g-vs-q-u} reproduces the $n$-dependence
of $g(k_{\rm F},{\bf q};\omega=0)$ 
in the large-$U$ limit. It agrees quite well with
the previous studies of Refs. \onlinecite{zeyher,kulic} and it shows the
the suppression of the large-${\bf q}$ scattering as the MIT is approaches
for $n \rightarrow 1$ and the corresponding development of
a small-${\bf q}$ momentum structure.
As a marginal difference we can note that the four-slave-boson approach
at finite $U$ predicts
a less pronounced peaked structure than the $U=\infty$ one slave-boson
method. This slight
discrepancy can be traced back to the different band narrowing factor between
our case ($Z_{\rm MF} \sim 2\delta/(1+\delta)$ as in Ref. \onlinecite{kotliar}
and the one-slave-boson $U=\infty$ treatment ($Z_{\rm MF} \sim \delta$).

One of the main results of our analysis is that it permits to investigate
the MIT and the momentum structure of the el-ph vertex
along a different path, namely increasing $U$ for {\em fixed} $n$.
As shown by the comparison between the three panels
of Fig. \ref{g-vs-q-u}, the predominance of forward scattering is
effectively suppressed as long as we move far from the MIT by reducing
the Hubbard repulsion $U$, in similar way with what is found as function
of $n$. This observation suggests that the small-${\bf q}$ structure
is only ruled by some macroscopic parameter which quantifies the
closeness to the metal-insulator transition, regardless the
microscopic parameters as $U$ or $n$. A natural candidate
along this perspective is the band narrowing $Z_{\rm MF}$, which
vanishes $Z_{\rm MF} \rightarrow 0$ at the MIT ($n=1$, $U\ge U_c$)
and $Z_{\rm MF} \rightarrow 1$ in the uncorrelated limit 
($n=0,2$ or $U=0$).

The effectiveness of the electronic correlation in giving rise to
a small-${\bf q}$ structure in the electron-phonon scattering
has been previously discussed
in literature in terms of a 
cut-off parameter $q_c$.\cite{gpsprl} On the physical ground
$q_c$ is expected to be related to the inverse of the correlation length $\xi$,
$q_c \propto 1/\xi$. At the MIT $\xi \rightarrow \infty$ and
$q_c \rightarrow 0$. For the practical purposes, it is often difficult to
extract a cut-off parameter $q_c$ from a complex structure as those shown
in Fig. \ref{g-vs-q-u}. To this aim we define a cut-off $q_c$ as the
full-width at the half-maximum (FWHM) of the electron-phonon vertex 
$g(k_{\rm F},{\bf q};\omega=0)$ along the $\Gamma$-$M$
line with respect its ${\bf q}=0$ value.
We would like to stress that this definition of $q_c$ is just one of the
possible choices and it does not pretend to be rigorous.
As matter of facts for systems far from the MIT, where no well defined
${\bf q}=0$ peak is recovered, no $q_c$ according such a definition
can be found. This should be interpreted as the absence of significant
small-${\bf q}$ modulations induced by the electronic correlation.
Note, on the other hand, that close to the MIT the small-${\bf q}$
peak can be always approximated as a Lorentzian, and the definition of
$q_c$ introduced above is indeed directly related to the inverse
of the correlation length $\xi$.

\begin{figure}
\centerline{\psfig{figure=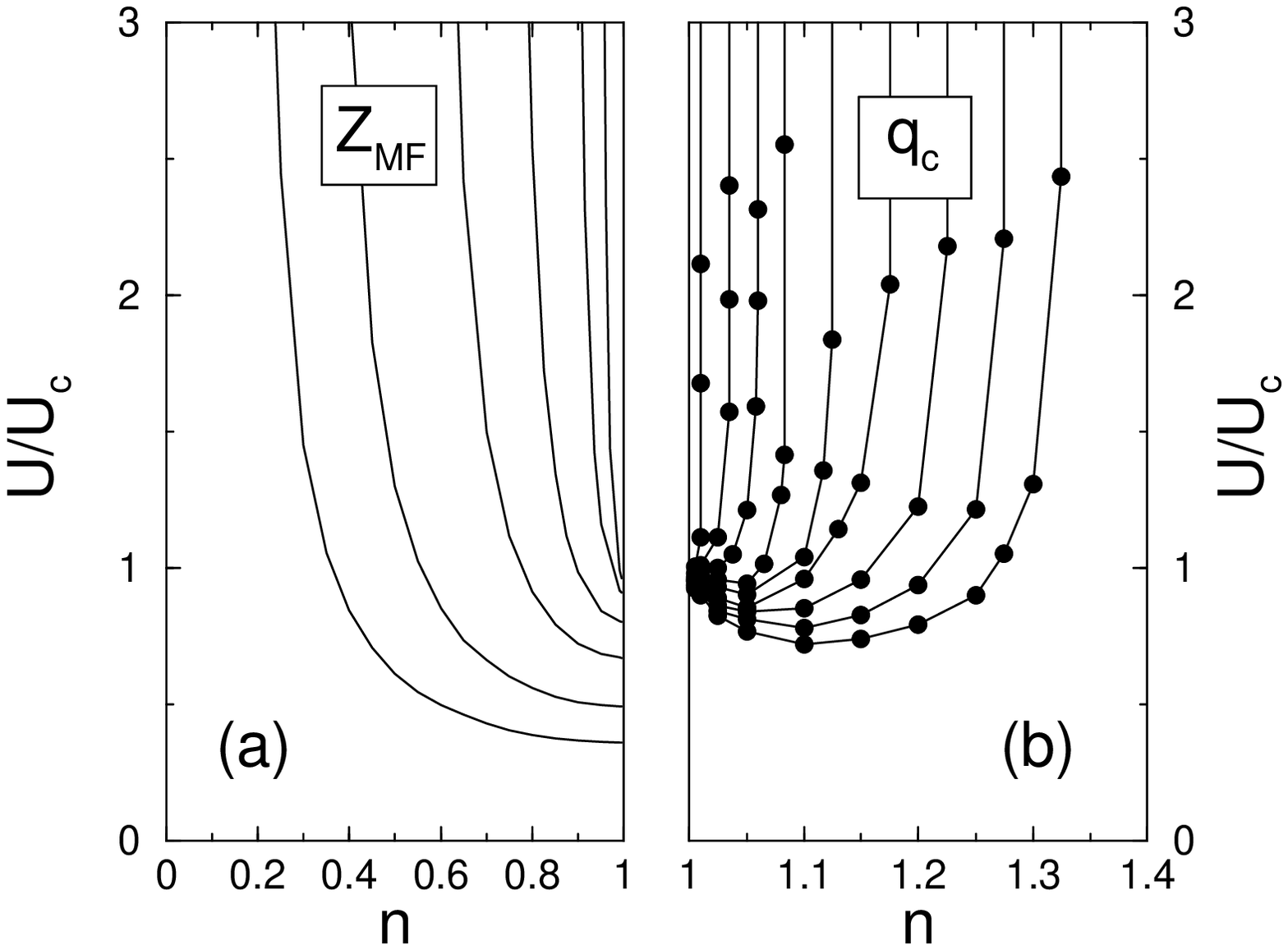,width=8cm,clip=}}
\centerline{\psfig{figure=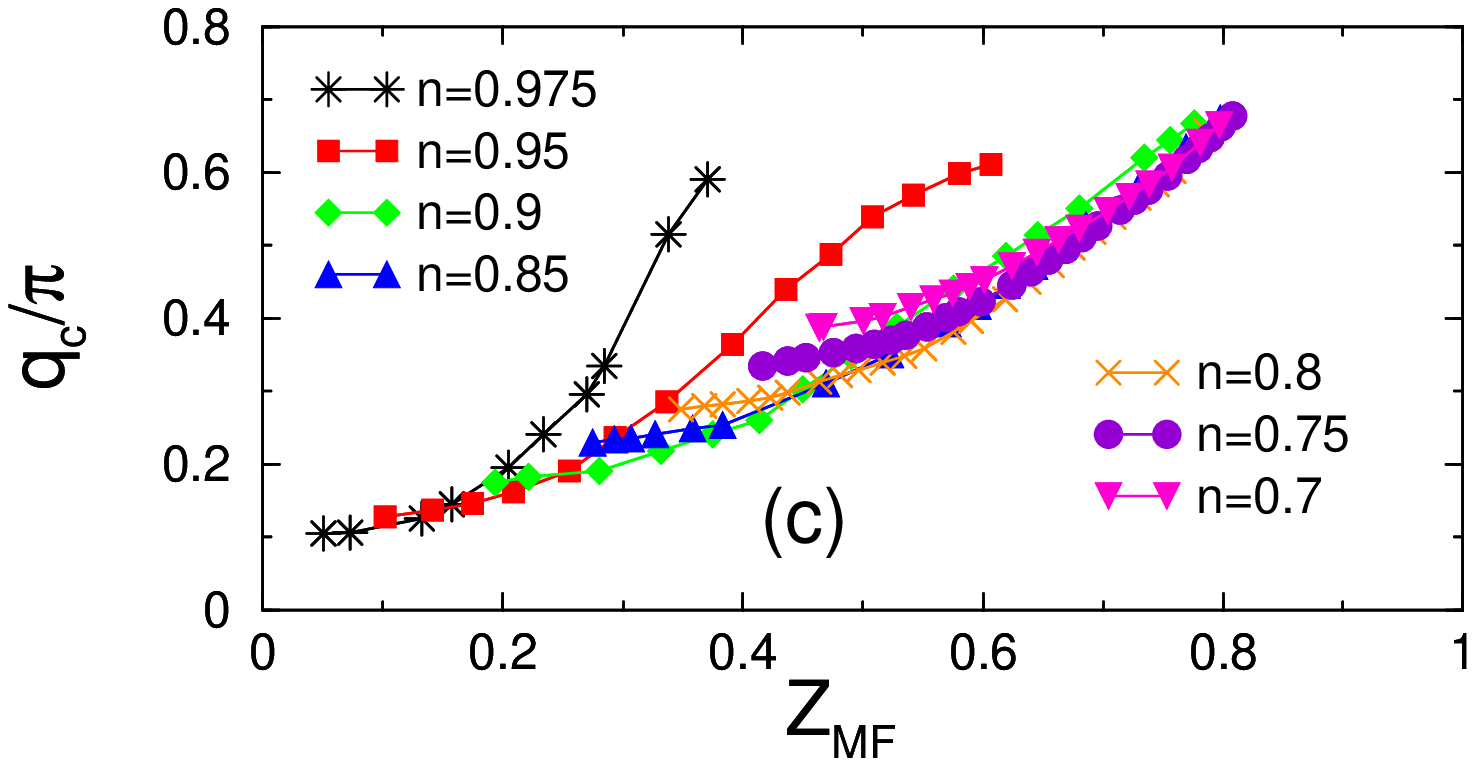,width=7cm,clip=}}
\caption{(color online)
Plot of the isolines for the quantities $Z_{\rm MF}$ (panel a)
and $q_c$ (panel b) in the $U$-$n$ space. Since both quantities are symmetric
for reflection with respect to the $n=2$ axis, they are shown respectively
in just one half-plane. Solid lines are (panel a, from right to the left):
$Z_{\rm F}=0.1, 0.2, 0.4, 0.6,0.8, 0.9$; (panel b, from left to right):
$q_c/\pi=0.2, 0.3, 0.4, 0.5, 0.6, 0.8, 1.0, 1.2, 1.4$. Panel (c): plot
of $q_c$ vs. $Z_{\rm MF}$ for different electron filling $n$.}
\label{phd}
\end{figure}

In Fig. \ref{phd}a,b we plot
the $Z_{\rm MF}$- and $q_c$-isolines
as obtained by applying our FWHM definition
in the whole $U$-$n$ phase diagram. The striking similarities between panels
(a) and (b) confirm that the small-${\bf q}$ structure is mainly ruled
by the strong band narrowing close to the MIT. This is pointed out in the most
remarkable way in panel (c) where the cut-off $q_c$, evaluated
for different $n$ and different $U$, is shown to lie on
an almost universal curve when plotted as function of $Z_{\rm MF}$.
This is a quite important result because the  band narrowing factor
$Z_{\rm MF}$, as well as quasi-particle spectral weight $Z_{\rm QP}$ which
is strongly related to $Z_{\rm MF}$, is in principle a physical quantity
which can be experimentally determined to provide an estimate of $q_c$.
The universality of $q_c$ as function of $Z_{\rm MF}$ points also towards
some general mechanism responsible for the small-${\rm q}$ scattering.
This mechanism has been discussed in literature in terms of proximity
to a phase separation which reflects in a charge response instability
at ${\bf q}=0$.
Before addressing the connection between small-${\bf q}$ scattering
and phase separation, we would note however
that, although the universal behaviour of $q_c$
vs. $Z_{\rm MF}$ appears quite robust,
some discrepancies appear for values of $n$ very close to $n=1$.
This is also in agreement with a comparison between
panels (a) and (b) which shows that for $n=1$,
in contrast to the $Z_{\rm MF}$ isolines, all the $q_c$-curves
converge to the $U=U_c$ point, where $q_c$ jumps from $q_c=\pi$ to $q_c=0$
from $U=U_c^-$ to $U=U_c^+$. A closer look at this discrepancy shows that
it can be related
to the presence of the Van Hove
singularity at $n=1$.
Indeed, while the quantity $Z_{\rm MF}$ does not depend on the
details of the electronic dispersion, 
in this case the ${\bf q}=0$ response functions
are described by an ideal metal with {\em infinite} density of states for any
$U \le U_c^-$.
It can be checked that the introduction of a finite $t'$ removes this
anomalous features and $q_c$ evolves in a smooth way increasing $U$ from
$U=0$ to $U=U_c$ for $n=1$.

As a final issue of this Letter we would like to comment about
the robustness of our results, based on a gaussian expansion within the
context of a finite-$U$ slave-boson formalism, with respect the inclusion
of higher order terms. A good check in this perspective would be
the comparison of our data with ${\bf q}$-resolved numerical calculations.
Unfortunately, simulation and numerical work along this line
is lacking in literature since the widely diffuse dynamical mean-field
approach mainly probes only local (${\bf }$-integrated) quantities
whereas exact diagonalization and QMC techniques are limited by 
finite size of the cluster and by finite temperature effects.
A seminal step to fill the gap has been provided by Huang {\em et al.}
who provided in Ref. \onlinecite{huang}, for the first time at our knowledge,
numerical support by QMC simulations
for the insurgence of small-${\bf q}$ structure in the
el-ph vertex function by increasing $U$.
As a by-product of their analysis, Huang {\em et al.} reported also
an interesting increasing of the absolute magnitude of the
el-ph vertex function for small momentum ${\bf q}=(\pi/4,\pi/4)$
in the large $U$ regime, whereas the increase
is absent for large momentum ${\bf q}=(\pi,\pi)$.
This is quite puzzling because the development
of a small-${\bf q}$ structure is thought to
be relative to the large momentum
scattering and it is usually not accompanied with
an increase of the magnitude of $g$ at ${\bf q}=0$.
The origin of this anomalous feature has been not well
understood so far and a possible connection with charge excitations
driven by the exchange term $J \propto t^2/U$ in the large $U$ limit
was mentioned.\cite{huang}

\begin{figure}
\centerline{\psfig{figure=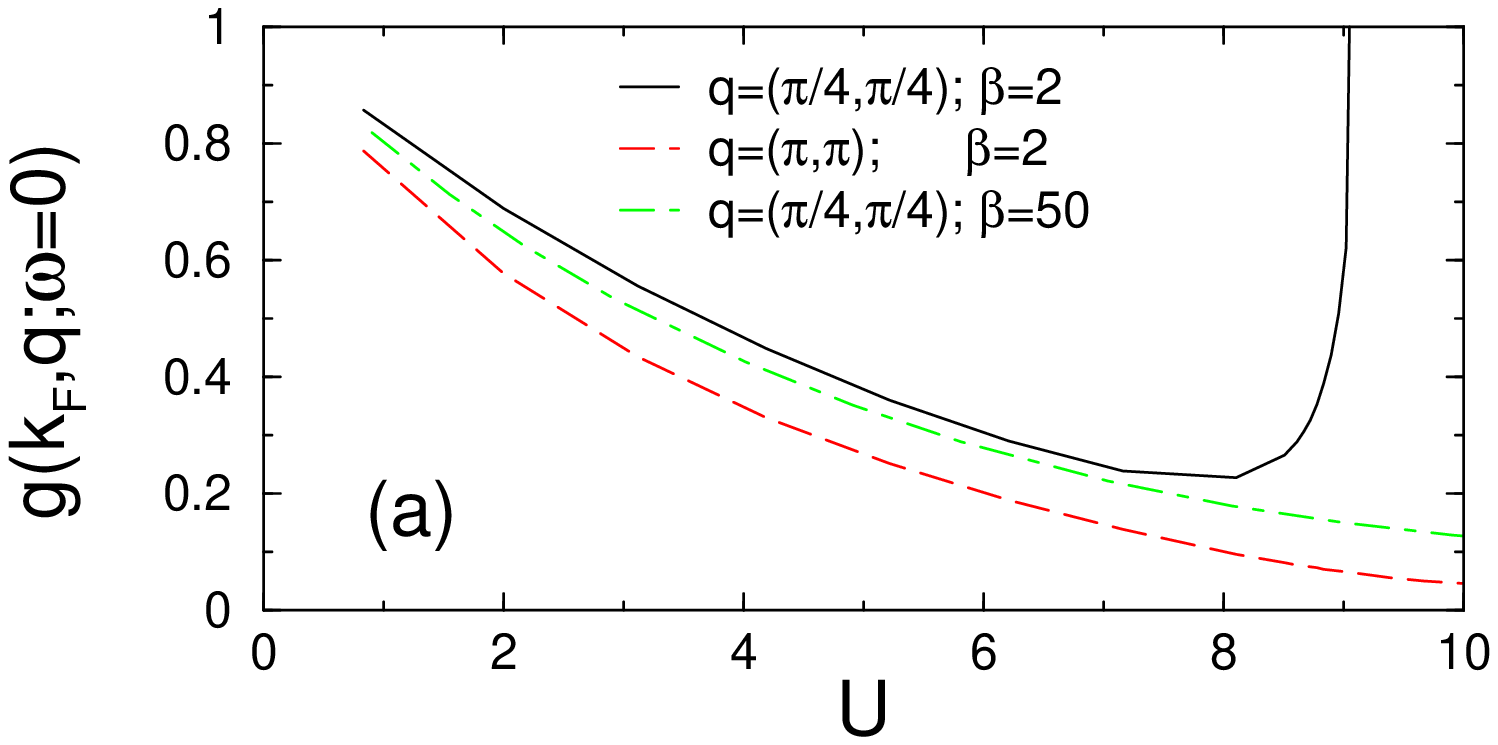,width=7.5cm,clip=}}
\centerline{\psfig{figure=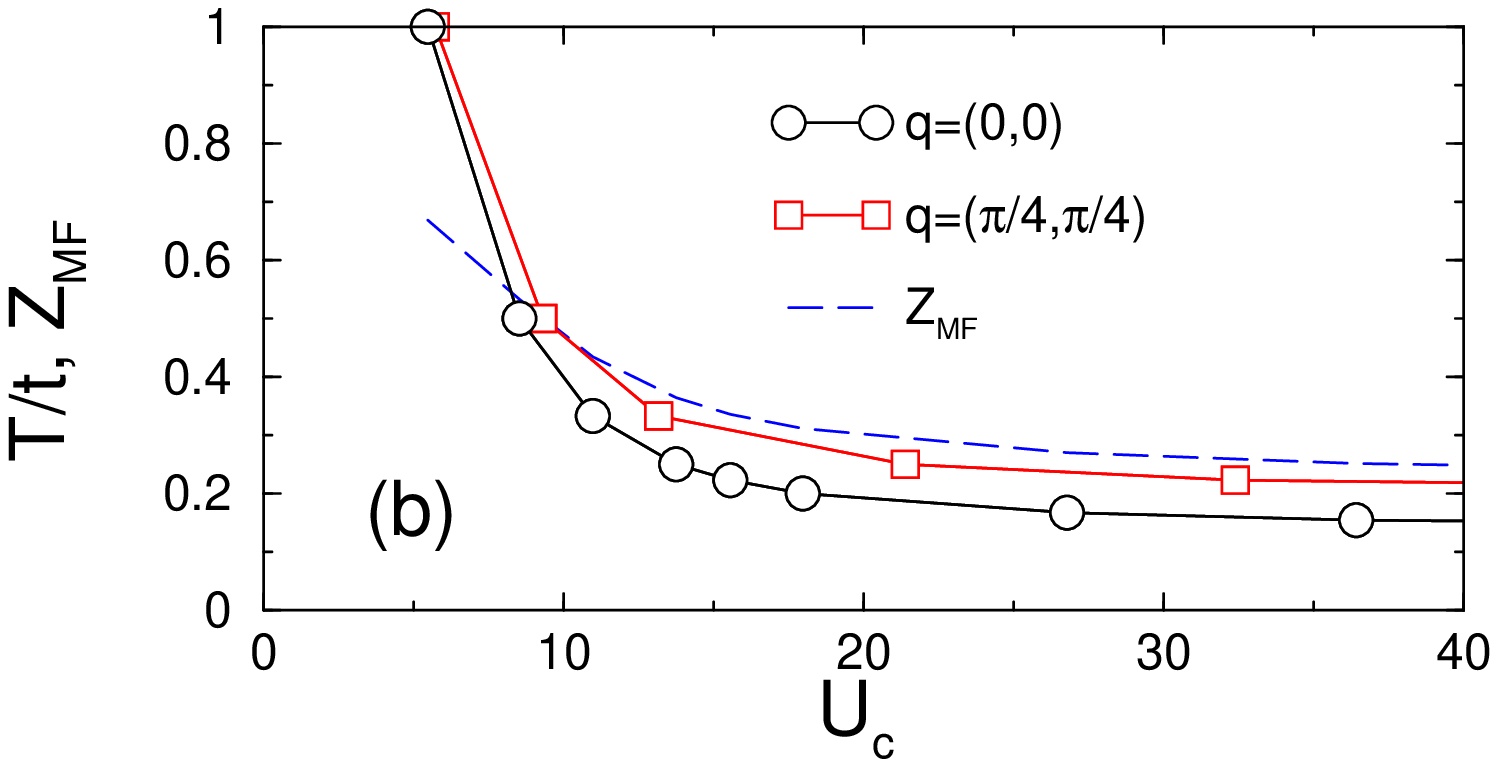,width=7.5cm,clip=}}
\caption{(color online) (a) Dependence of the el-ph vertex function on $U$ for
$n=0.88$ and for ${\bf q}=(\pi/4,\pi/4)$ (solid line, $\beta=2$),
${\bf q}=(\pi,\pi)$ (dashed line, $\beta=2$), and
${\bf q}=(\pi/4,\pi/4)$ (dot-dashed line, $\beta=50$).
(b) Phase diagram for phase separation (${\bf q}=0$) and
charge instability (${\bf q}=\pi/4,\pi/4$). The dashed line shows the
factor $Z_{\rm MF}$ evaluated along the phase separation instability line}
\label{f-huang}
\end{figure}

In order to test to reliability of our results we have addressed this
issue within the context of our slave-boson formalism.
In Fig. \ref{f-huang}a we show the dependence of the el-ph
vertex function $g(k_{\rm F},{\bf q};\omega=0)$ on the Hubbard
repulsion $U$ as calculated
within our approach for the same parameters as in Ref. \onlinecite{huang},
namely $n=0.88$, ${\bf q}=(\pi/4,\pi/4)$, ${\bf q}=(\pi,\pi)$,
$\beta = 2$, where $\beta$ is the inverse of the temperature $T$
in units of the nearest-neighbor hopping element $t$.
The agreement
between our findings and the data of Ref. \onlinecite{huang}
is quite excellent,
although quantitative differences are of course present, In particular we
find that, while for large momenta $g(k_{\rm F},{\bf q};\omega=0)$
steadily decreases with $U$, for ${\bf q}=(\pi/4,\pi/4)$
it shows
an upturn for $U \simeq 8$ until a divergence occurs for
$U_c \simeq 9.3$, signalizing an
charge instability.\cite{ccp} According this view one is attempted
to associate the upturn of the el-ph vertex function
$g(k_{\rm F},{\bf q};\omega=0)$ 
as function of $U$ reported in Ref. \onlinecite{huang} 
as an incipient transition
towards some charge instability or phase separation.\cite{notehuang}
This does not imply however that phase separation is effectively established,
and it should be remarked that the actual occurrence of phase separation
in the Hubbard model is still object of debate.\cite{cosentini}
In particular the large temperature $\beta=2$ at which the
QMC were performed could play an crucial role in that. To investigate
this final issue we evaluated again the el-ph vertex function for the same
parameters but a much lower temperature $\beta=50$.
As shown in Fig.
\ref{f-huang}a, quite surprising the charge instability disappears
decreasing the temperature.

In order to understand in more detail the origin
of the phase separation instability as function
of the temperature $T$ we show in Fig. \ref{f-huang}b
the phase diagram in the $T$ vs. $U$ space with
respect to phase separation (${\bf q}=0$) and
to charge ordering (${\bf q}=(\pi /4,\pi /4)$).
Phase separation as well as charge instabilities occur
in our slave-boson calculations only above a certain temperature
$T/t \bsim 0.2$ ($\beta \lsim 5$).
A interesting insight comes from the comparison of the critical
temperature $T_c$ at which the instability towards phase
separation occurs with the band narrowing factor $Z_{\rm MF}$ 
(Fig. \ref{f-huang}).
The similar dependence of $T_c$ and $Z_{\rm MF}$ on $U$
points out
that the onset of phase separation by increasing temperature
is ruled by the comparison between the temperature $T$ and
the effective bandwidth $W=Z_{\rm MF} 8t$ energy scales.
In particular phase separation is established when $T$ becomes
of the same order of $Z_{\rm MF} t$.
We would like to stress that in principle the
phase separation instability found by
our slave-boson calculations
could be washed out
when higher order fluctuations than the gaussian ones are taken into account.
The agreement between our data and QMC results suggests however
that the correlation between small-${\bf q}$ modulation,
phase separation and the upturn as function of $U$
could be a robust feature of correlated systems. 
At a speculative level, we think that
the small coherent part of the
electronic spectral weight (with the electronic dispersion
scaling with $Z_{\rm MF}$) could be responsible for the tendency
towards phase separation, which is however prevent by the presence of the
incoherent states. Further analytical and numerical
work on this direction will help to shed new light on this issue.


\begin{thebibliography}{99}
\bibitem{notehuang}
Investigation of a
phase separation instability was prevented
in Ref. \cite{huang}
since the momentum discretization due to the
finite cluster.
\end{thebibliography}
\end{document}